# Analysis of the impact of studies published by Internext - Revista Eletrônica de Negócios Internacionais

Carlos Luis González-Valiente[1]

**ABSTRACT**

This paper presents a citation analysis of Internext-Review of International Business to detect the impact caused by papers published for the period 2006-2013. The Publish or Perish (PoP) software is used, which retrieves articles and citations from Google Scholar database. As part of the applied indicators are: the distribution of authors by articles, citations per year, citation vs. self-citation, journal's citable vs. non citable documents, journal's cited vs. uncited documents, co-word analysis, and H Index. A total of 131 articles were obtained for 153 citations made until June, 2014. Most articles present multiple authorship. It is also detected an ascending line in the citation. The Journal has very low levels of self-citation, showing that most citing sources are Brazilian journals. The most cited articles have been published in the early years (2006-2008); whose main topics are related with the internationalization theory and strategy, the transaction analysis, and the corporate governance. The Internext' H Index is 6 and the G Index is 9.

## 1. Introduction

One of the most common ways to identify the impact caused by studies in the scientific community is through citation analysis. Marx & Bornmann (2013) argue that citations can be used as a way to measure the development of research, and its impact is one of the aspects that determine the quality of a journal's articles, complemented by its accuracy and originality. Such investigations reveal the level of quality of different journals and the extent of their influence on one or more disciplines. Based on other perspectives, external factors that have a bearing on the impact of journals have also been explored (YUE & WILSON, 2004).

In studying the impact of journals, although it varies from one to another (MOED, 2005; LEYDESDORFF, 2008), one should not solely focus on the more prestigious journals, but also on those which are somewhat "peripheral" in nature. An excellent means for determining this impact has been the Journal Citation Report (JCR) of the Institute of Scientific Information (ISI), but this only works for journals indexed by this system. It is therefore difficult to identify the impact of journals not included by the ISI, which is a source of frustration for many editors whose journals are not recognized by this major platform (LEYDESDORFF, 2008).

A database which has democratized impact analysis is Google Scholar, since it includes a wide variety of scientific material that can be uploaded to institutional repositories by authors, editors or other collaborators[2]. Google Scholar, through the use of mathematical algorithms, generates statistics for citations of articles, authors and journals, making it possible to calculate and show the h-index and citations received. These capabilities make it very comparable to other information resources such as Web of Knowledge (WoK) and SCOPUS, precisely due to its tools for analysis and measurement of the activity and scientific results (AGUILLO, 2012).

---

[1] Grupo Empresarial de la Industria Sidero Mecánica, Habana, Cuba. Corresponding author: Email: carlos.valiente@fcom.uh.cu or carlos.valiente89@gmail.com

[2] http://scholar.google.com/

Although many authors have outlined the advantages and disadvantages of Google Scholar for quantitative and qualitative studies in science and technology, it is evident that it provides a new way to evaluate journals. This study focuses on the case of Internext, a journal in the area of Management and International Business. The objective of this article is to perform a citation analysis of Internext-Revista Eletrônica de Negócios Internacionais, in order to identify the impact caused by studies published therein.

## 2. Literature review

### 2.1 Internext - Revista Eletrônica de Negócios Internacionais

Internext is a journal that is scientific in nature, published by the Master's and PhD Program in International Management of the School of Advertising and Marketing (ESPM) in Brazil. Its first issue was in 2006. Until 2013, it came out twice a year, and after that, three times a year. Its thematic focus, as mentioned earlier, is Management and International Business, and it seeks to delve into different branches of management, such as: strategy, finance, people management, organizational studies and others.

The sections designated for publications of authors contain articles, abstracts, editorials, essays and teaching cases. With the exception of editorials, all these sections are peer reviewed. Once published, they are indexed by databases, directories and repositories, such as: Diretório de Políticas de Acesso Aberto das Revistas Científicas Brasileiras (DIADORIM), Directory of Open Access (DOAJ), EBSCO Publishing-Fonte Acadêmica, Gale Cengage Learning- Academic OneFile, Google Scholar, LATINDEX, Portal de Periódicos da CAPES, Scientific Periodicals Electronic Library (SPELL), Sumários de Revistas Brasileiras, and Sistema Eletrônico de Editoração de Revistas (SEER). Its classification, according to CAPES' Integrated System[3], is3 B2.

### 2.2 Citation analysis of management and international business studies

It is not uncommon in bibliometric characterizations to find themes associated with research on Management and International Business (MaIB). In this regard, there are many cases that can be cited. For example, Chan Fung & Leung (2006), working with ten years of publications (1995-2004) from leading international journals, examined global patterns in international business and the academic ranking of these journals. Similarly, Griffith, Tamer & Xu (2008), who focused on six leading journals in the area of MaIB, detected emerging themes that were explored from 1996 to 2006. This served to demonstrate the consistent progress made on the topic, which would help with the development of future studies, as the authors point out. On the other hand, Harzing (2008) investigated the impact that MaIB journals – indexed by JCR from 2006 to 2008 – have had on the international community. Among the most significant results, the following titles were grouped in descending order according to h-index: Journal of International Business Studies, Journal of World Business, International Business Review, Management International Review, Journal of International Management, International Studies of Management & Organization, and Thunderbird International Business Review.

From a somewhat different perspective, De Moura Carpes et al. did a citation analysis based on journals indexed by Web of Science (WoS) and determined the domains of productivity of the authors in relation to International Business, from 1997 to 2010. In turn, Reis et al. (2013), using 3,639 articles retrieved from ISI-Wok, performed a citation analysis to identify the domains of knowledge researchers relied on for their studies in the area of International Business. It provided the most cited authors and works, as well as an extensive co-citation network, based on 20 articles that had caused the greatest impact.

---

[3] http://qualis.capes.gov.br/webqualis/principal.seam

Taking advantage of another type of source, Hofer et al. (2010) used presentations at conferences of the Academy of International Business (AIB) to create the structure of research interests involving International Business studies, based on a co-word analysis, in which core and peripheral topics were highlighted. It is important to note that the area of International Business is not the only core theme to have been evaluated, but other subjects related to it have also been the object of bibliometric interest, such as international management (WERNER, 2002), Resource-Based View[4] (PORTUGAL, RIBEIRO & KRAMER, 2011), business studies (LUOR et al., 2013), and others As mentioned earlier, most authors use the most important Management and International Business journals for their research – all from major sources that facilitate citation analysis, such as JCR, WoS and WoK, alike. Many note the prerogatives these databases offer, since they are very elitist for the inclusion of journals, and consequently limit themselves to few sources for citation analysis and evaluate only the active period of inclusion of journals in those resources. A study which democratizes results was done by Harzing & Van Der Wal (2009), where Google Scholar was used as the reference source to measure the impact of journals in the field of Economics and Business, through an analysis of the h-index and comparing it with the ISI Journal Impact Factor, from 2000 to 2005.

Lastly, albeit not so significantly, are the results of research in which journals were studied independently; such as the Journal of International Business Studies (JIBS). For example, Inkpen & Beamish (1994) assessed its 25 years of publications starting from 1970. Chandy & Williams (1994), for the period from 1984 to 1993, and Phene & Guisinger (1998), for eleven of its publication years (1981-1991), also focused on the productivity and growth of citations of the JIBS. A little closer to the Brazilian context is the study by Tinoco (2005), which analyzed the citation patterns of the country's leading management journals: Revista de Administração Contemporânea (RAC); Revista de Administração de Empresas (RAE), Revista de Administração Pública (RAP) and Revista de Administração (RAUSP), from 1997 to 2002. Another article is the one by Machado-da-Silva et al. (2008), which also studied the impact of management journals (21 in this case), covering three years (2005-2007). In another study, Wood Jr. & Vouga (2008) presented a ranking of the scientific production from Business Administration programs in Brazil, using the scientific articles published in the major journals of that country from 2002 to 2006.

Coincidentally, many of these journals were also used by Tinoco (2005), Machado-da-Silva et al. (2008) and Da Silva (2012). As a specific case of journals evaluated independently, there is the study by Francisco (2011), which focused on the business management journal Revista de Administração de Empresas (RAE).

3. Methodology

Document analysis was used in this study to gather information on the object of study, which enabled an approach related to the antecedents of the categories examined here. However, the most solid perspective applied entails bibliometrics. The population of this study consists of the articles published by Internext and works which cite these articles. Therefore, the unit of analysis is composed of published articles as well as citing articles. The variables or categories analyzed therein are the following:

- Published articles: citations, year of publication, authors, words from the title and keywords.
- Citing articles: year of publication, type of source.

---

[4] Term commonly used in English.

From there, the following indicators were generated, which were also adapted from those developed by SCImago Journal & Country Rank[5]:

- Distribution of authors per article. Number of authors who collaborate according to each article published by the journal.
- Citations per year. Number of citations the articles received according to the years in which these citations were made.
- Citation versus self-citation. Evolution of the total number of citations received versus the number of self-citations made by the journal itself during the period covered.
- Citable journals versus non-citable documents. Citable documents are considered as those that have been published by scientific journals. This indicator, therefore, shows the distribution of the total number of citations received by journals (primary sources) versus citations received from other sources of a secondary nature.
- Citation of the journal versus uncited documents. Distribution by publication years of the number of articles that were cited, at least once, as opposed to the total number of articles that were not cited.
- Co-word analysis. Identification of words that co-occurred two or more times from the main categories extracted from the titles and keywords of those articles that were cited at least once.
- H-index or Hirsch Index. Its calculation is purely automatic and is well described in the PoP user manual.

The core tool used was the software Publish or Perish (PoP), developed by Anne-Wil Harzing (2007). This program retrieves and analyzes academic citations from the Google Scholar database[6]. It states the set of indicators to be measured and, once the search strategy has been formulated, pulls up results which, through a graphical interface, can be copied or saved in different formats. This software has also been used in other studies that produced important findings related to citations given on multiple topics and sources (e.g.: ARENCIBIA, 2008; Harzing & VAN DER WAL. 2009; JACSÓ, 2009).

Since the object of study was the journal Internext, this term was entered into the search box of the PoP Journal Impact section. Every article published between 2006 and 2014 was obtained, of which duplicates and those from 2014 were eliminated, since 2014 was still underway at the time of conducting this study. In the end, there were 131 articles, with a total of 153 citations up until June 2014 (see Figure 1 in Appendix 1).

---

[5] http://www.scimagojr.com/index.php
[6] http://scholar.google.com/

Besides the facilities provided by PoP, the tool EndNote X4 was used to create two libraries, one with the references of published articles and the other with those from citing articles. This normalized each of the variables to be analyzed, as well as eliminated duplicate records. For the co-word analysis, Bibexel software[7] was used, while VOSviewer 1.4.0[8] was used to view the word relationships through bibliometric maps.

The measurement of the data from the population sampled was done by counting, with the support of Microsoft Excel. However, many measurements were automatically performed by PoP itself, such as number of articles, number of citations and citations per article, number of articles per author, h-index and g-index.

## 4. Results and discussion

As mentioned in the methodology section, a total of 131 articles were retrieved, corresponding to a period of eight very active and uninterrupted years. First, it is worth noting that one of the aspects that modern science, as well as the area of scientific production, has been taking into account is multiple as opposed to individual authorship. In the case of Internext, most of its articles were written by two authors, with a lower number of articles written by four or five authors (see Table 1).

**Tab.1** Distribution of authors per articles

| Total of authors | Total of articles |
|---|---|
| 1 | 35 |
| 2 | 44 |
| 3 | 35 |
| 4 | 15 |
| 5 | 2 |

**Source:** PoP, 2014

As seen in Figure 1, the level of citations received by Internext has been rising considerably, totaling 153. One year after the launch of the journal, its articles started being cited, with the highest figures in 2011 to 2013. The analysis also included the first half of 2014, up until which time 14 citations had been made.

---

[7] www8.umu.se/inforsk/Bibexcel
[8] www.vosviewer.com

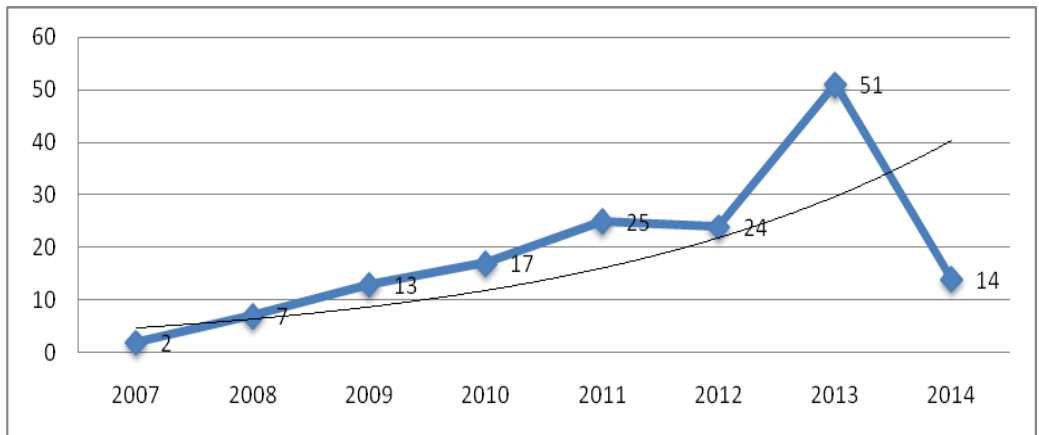

Figure 1. Distribution of citations per year
Source: Prepared by the author using PoP

According to Figure 2, the journal's self-citation rate (13.72%) is notably low. These self-citations are unevenly distributed and only occur in 2010, 2013 and 2014. This phenomenon of self-citation by journals has a bearing on the manipulation of their impact factor (ENGQVIST & FROMMEN, 2008; FRANDSEN, 2007); hence, there has been ongoing discussion on whether or not to include these citations in the generation of h- and g-indexes (COSTAS, VAN LEEUWEN & BORDONS, 2010).

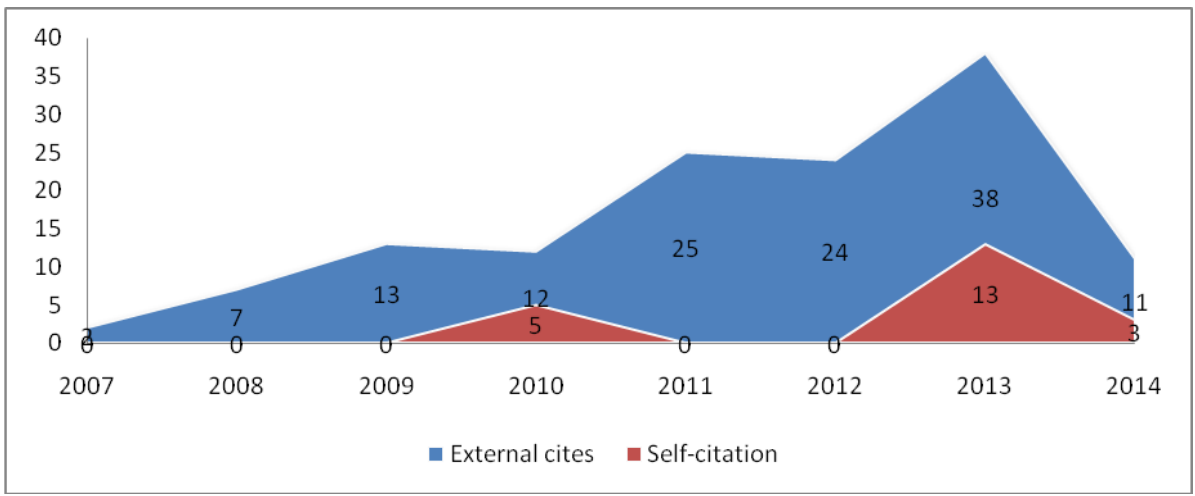

Figure 2: Levels of citation and self-citation in relation to Internext
Source: Prepared by the author using PoP

Lawani (1982) classifies self-citation into two types: synchronic and diachronic. In the case of this study, the results are diachronic since the calculation performed was based on the total number of self-citations in relation to the total number of citations Internext has received. As far as the type of sources that cite Internext, it can be seen in Figure 3, the interest of citable sources (52.28%) has been higher than non-citable ones (47.71%). The citations from citable sources come from over 43 scientific journals, including: Internext-Revista Eletrônica de Negócios Internacionais da ESPM (13.72%), Informações económicas (3.92%), RAM. Revista de Administração Mackenzie (1.96%), BASE-Revista de Administração e Contabilidade da Unisinos (1.96%) and Revista Ibero-Americana de Estratégia (1.96%). Of the journals, 83.72% are from Brazil and the rest from other countries, such as the United States (4.65%), Argentina (4.65%), Chile (2.32%), Spain (2.32%) and England (2.32%). Non-citable sources include theses (36.64%) and conference presentations (9.8%). The high presence of citing theses is due to the fact that many academic programs, mostly for Master's degrees, have repositories which enable the representation and retrieval of their references. This, like other document types, is indexed by Google Scholar, a factor that has influenced the democratization of the impact of research, when making bibliometric assessments.

It is noticeable that the impact caused by Internext is not international, which may due, to a certain extent, to language – hence the large concentration of citations from within Brazil itself. If considered from this angle, it is important to note that 90.07% of the articles are published in Portuguese, and the rest in English (9.16%) and Spanish (0.76%), whereas for citing articles, there are 69.28% in Portuguese, 14.37% in English and 1.96% in Spanish.

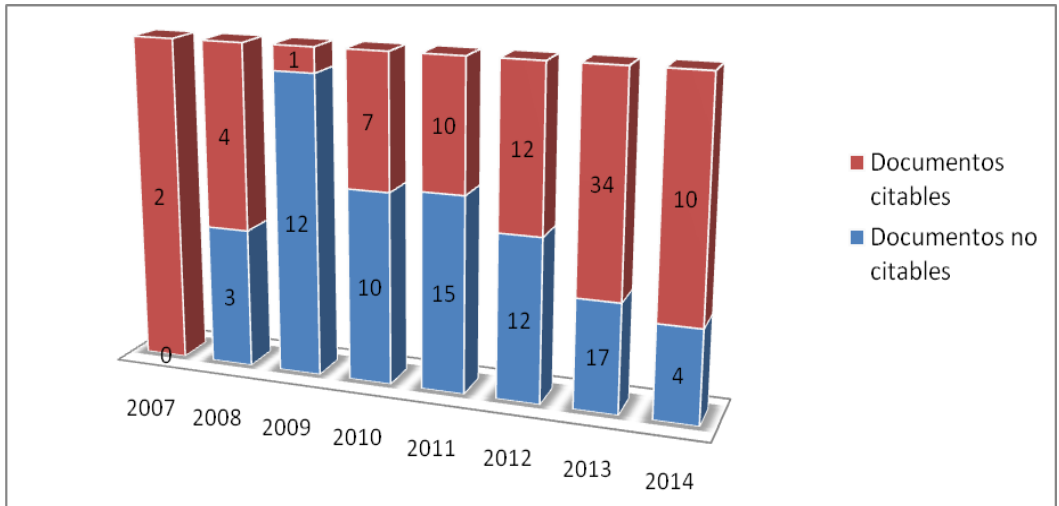

**Figure 3**: Citable and non-citable sources that cite Internext
**Source**: Prepared by the author using PoP

Focusing on another indicator, it was detected that Internext tends to publish an average of 16.37 articles per year (see Figure 4). Of these, the number of uncited documents is higher than cited ones. On average, 4.5 articles tend to cause an impact on cited documents each year. It can also be seen in Figure 4 that the number of articles cited per issue has fallen; but if this trend is correlated with the one reflected in Figure 1, an imbalance coexists in the growth lines in relation to the years in which the citations were made.

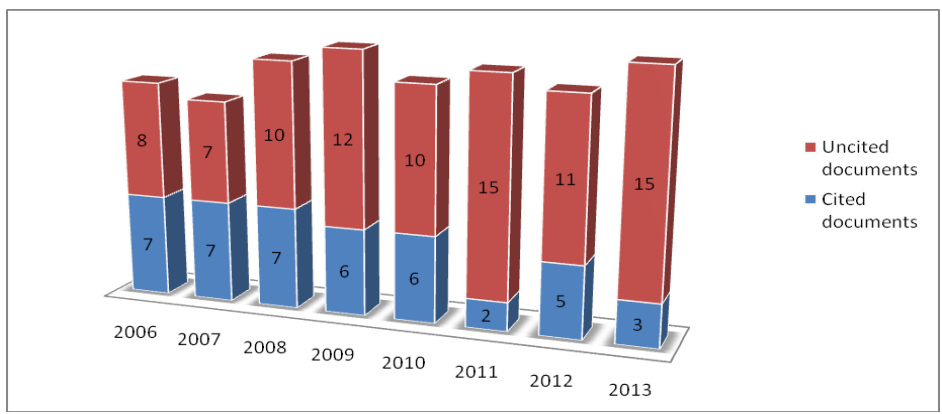

**Figure 4**: Behavior of the number of cited articles according to year of publication
**Source**: Prepared by the author using PoP

One factor that may have a bearing on the decreased number of articles cited per year is the fact that the strategy of Internext to publish a larger number of articles per year, which results in a very notable difference in the number of articles cited until now. Furthermore, another key element to highlight may be the delay in the publication of citing articles, an issue that is implicit in any process of scientific communication. Table 2 shows which titles are the most cited ones in the journal, specifically those which have received five or more citations. These studies account for 50.32% of the total citations, demonstrating that the articles that caused the greatest impact were published during the early years of Internext (2006-2008).

**Tab.2** Most cited articles from the journal Internext

| Author(s) | Title | # of cites | % of cites |
|---|---|---|---|
| Jorge Manuel Teixeira Carneiro; Luis Antônio Dib | Avaliação comparativa do escopo descritivo e explanatório dos principais modelos de internacionalização de empresas; v. 2, n.1, p.1-25, 2007. | 33 | 21.56% |
| Geni Satiko Sato | Vinhos brasileiros: é possível a internacionalização; v. 1, n. 1, p. 243-259, 2006. | 10 | 6.53% |
| Walter F. de Moraes; Brigitte Renata Bezerra de Oliveira; Érica Piros Kovacs | Teorias de internacionalização e aplicação em países emergentes: uma análise crítica; v. 1, n. 1, p. 221-242, 2006. | 9 | 5.88% |
| Luciana Florêncio de Almeida; Décio Zylbersztajn | Crédito Agrícola no Brasil: uma perspectiva institucional sobre a evolução dos contratos; v. 3, n. 2, p. 267-287, 2008. | 8 | 5.22% |
| Nadia Wacila Hanania Vianna; Sheila Regina Almeida | A decisão de internacionalizar; v. 6, n. 2, p. 1-21, 2011. | 6 | 3.92% |
| George Albin R. de Andrade | Estudo econométrico dos efeitos da migração para OIGC: índice de ações com governança corporativa diferenciada da Bovespa; v. 3, n. 1, p. 39-53; 2008. | 6 | 3.92% |
| Leonardo Nelmi Trevisan | Pressão cambial ea decisão de internecionalização: o caso Marcopolo no período 2004/2005; v. 1, n. 1, p. 203-220, 2006. | 5 | 3.26% |

**Source:** Prepared by the author using PoP

Through a content analysis of these articles, it was found that the most relevant topics were related to studies on internationalization theory (TEIXEIRA & DIB, 2007; DE MORAES, BEZERRA & PIROS, 2006) and internationalization strategies (SATIKO, 2006; VIANNA & ALMEIDA, 2011; (TREVISAN, 2006); transaction analysis (ALMEIDA & ZYLBERSZTAJN, 2008) and corporate governance (ANDRADE, 2008).

This shows that, from the perspective of cited articles, there is a high level of specialization in the journal in relation to its thematic scope. These topics have also been addressed considerably in preliminary studies, as well as more comprehensive ones, such as those by De Moura et al. (2010), Hofer (2010) and Werner (2002).

An explicit analysis of the most co-occurring terms in the 43 cited articles, found that there was co-occurrence in 21 of the 109 total categories. The most representative in terms of frequency of appearance (fa) were internationalization (fa: 16), strategy (fa: 5), internationalization strategy (fa: 5), market (fa: 5), company (fa: 4), brand (fa: 3), international business (fa: 3) and knowledge transfer (fa: 3). The co-occurrence map in Figure 5 shows five main thematic clusters, although cluster 5 is not very representative due to the small grouping of terms it has. The strongest relationships between keywords are from the categories company-internationalization (fr: 2), internationalization strategy-resource-based view (fr: 2), direct foreign investment-market (fr: 2), and company-strategy (fr: 2). The main thematic lines that have been the object of citation of articles from Internext are consolidated on these relationships, which is no different from the results obtained from other research (MELO & ANDREASSI, 2010).

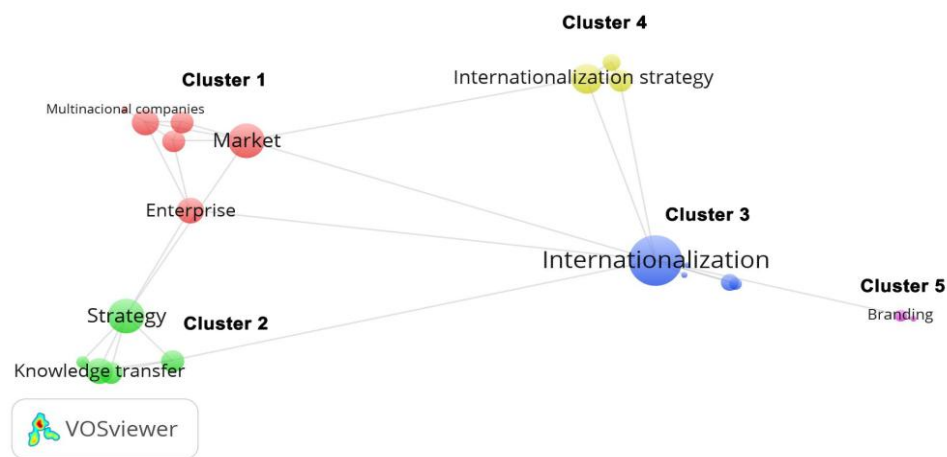

**Figure 5**: Map of co-occurrence of terms
**Source**: VOSviewer 1.6.1

Moving on to the last indicator, it was determined, through PoP, that the h-index of Internext is 6. This software also indicates that its g-index is 9, while its contemporary h-index is also 6. A comparative analysis was also performed between Internext and all impact journals from Brazil. These journals were selected through SCImago Journal & Country Rank by means of a search targeting the generic area of Business, Management and Accounting in Brazil, for 2012. These results are presented in Table 3, where it can be seen that Internext is in last place among six other journals.

**Tab. 3:** Comparative analysis of the h-index of Internext with similar journals from Brazil

| Journal | H Index (Google Scholar)[9] | H Index (SCImago Journal & Country Rank) |
|---|---|---|
| RAE. Revista de Administracão de Empresas | 57 | 2 |
| Gestão e Produção | 34 | 5 |

---
[9] http://scholar.google.com.cu/citations?view_op=top_venues&hl=es

| | | |
|---|---|---|
| Revista Brasileira de Orientação Profissional | 22 | 1 |
| Brazilian Administration Review | 14 | 3 |
| Revista Brasileira de Gestão de Negocios | 11 | 3 |
| InternexT-Revista Eletrônica de Negócios Internacionais da ESPM | 6 | - |

**Source:** Prepared by the author using PoP and data from the Internet

Multiple factors may have a bearing on Internext did not manifest balanced results compared to other journals, such as: visibility in impact databases, coverage in years, language and relevance of the studies, among others. As shown, the h-indexes, according to SCImago Journal & Country Rank, are not very high, which is not parallel to those of Google Scholar. In terms of projections, Internext should focus on other levels and strategies because its h-index lags far behind those of the main Brazilian journals.

## 5. Final considerations

The main objective of this study was to present approaches regarding the extent of the impact caused by articles published by Internext within the scientific community. Although results are presented comparing it with the leading journals in Brazil, the number of citations has been growing. However, the highest rate of citations is related to its first issues and only by journals that are its counterpart in Brazil.

Therefore, future projections could be aimed at the internationalization of the contents, beyond the barriers that may coexist. Even though the accuracy of all the data cannot be confirmed, at least the results serve as a platform for redesigning or proposing new strategies for the publication of its materials. To increase the citations that Internext could receive, one suggestion would be indexation in impact or high profile international databases, such as ISI or Elsevier; among which figure WoS, JCR, Scopus and others. Another alternative could be to publish studies addressing issues that are not so local in nature and that contribute toward the generic development of the discipline to which the journal is directed. Translation into Portuguese of leading articles in the field that have already been published in mainstream journals, as well as the publication of studies in English and Spanish, are another two approaches that could enhance the journal's visibility.

Another bibliometric analysis that could be done on Internext in the future is related to indicators based on analysis of publications; based on such analyses it could be determined which thematic lines are addressed the most, which authors contribute the most, the levels of collaboration by institutions and countries, and other related information.

Although it would likewise be relevant to perform a citation analysis that is synchronic in nature, i.e., that analyzes the citations made in articles to determine the intellectual base that characterizes the knowledge of authors who publish in the journal.

The bibliometric perspectives may be many, but all the results are directed toward the same end: making decisions about scientific activities; decision-making that is not only in the hands of publishers but also editorial boards, reviewers, authors, and even consumers. Once again, it is worth noting the opportunity provided by the Google Scholar database, which, from a democratic perspective, offers all scientific journals the possibility of a metric analysis.

## 7. Appendix

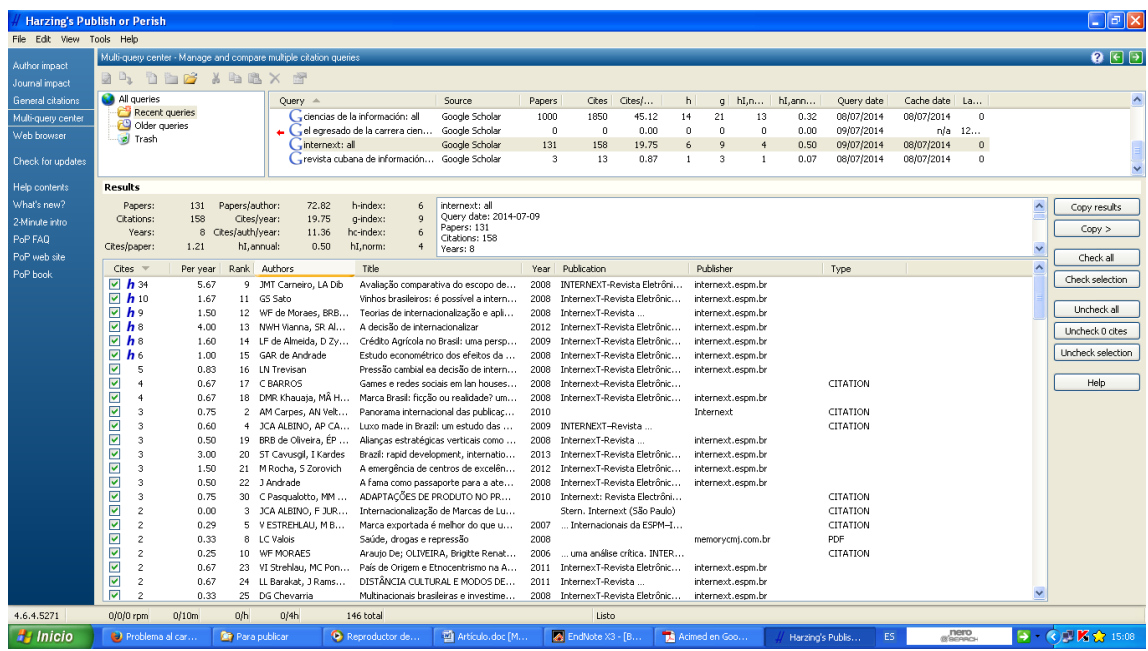

**Figure 1:** Interface of the results obtained by PoP regarding the impact of the journal Internext from 2006-2013
**Source:** PoP, 2014

Análisis del impacto de las investigaciones publicadas por Internext - Revista Eletrônica de Negócios Internacionais

Carlos Luis González-Valiente

Grupo Empresarial de la Industria Sidero Mecánica, Habana, Cuba

RESUMEN

Se presenta un análisis de citas de Internext - Revista Eletrônica de Negócios Internacionais, para detectar el impacto que han causado las investigaciones publicadas durante el periodo 2006-2013. Se utiliza el software Publish or Perish (PoP), el cual recupera los artículos y las citas desde la base de datos Google Scholar. Como parte de los indicadores aplicados se destacan: la distribución de autores por artículos, las citas por año, la citación vs. autocitación, las revistas citables vs. documentos no citables, la citación a la revista vs. documentos no citados, el análisis de co-ocurrencia de términos y el Índice H. Se obtuvieron 131 artículos para un total de 153 citas recibidas hasta junio de 2014. Los artículos tienden a presentar autoría múltiple, detectándose una línea ascendente en la citación. Han sido muy bajos los niveles de autocitación, evidenciándose que la mayoría de las fuentes citantes son revistas brasileñas. Han sido más los artículos no citados que los citados, causando mayor impacto los publicados durante los primeros años (2006-2008). Los artículos más citados guardan relación con las temáticas de teoría y estrategia de internacionalización, análisis de transacciones y gobernanza corporativa. El Índice H según PoP es de 6 y el Índice G es de 9.

**Palabras-clave:** Bibliometría; Impacto de revistas; Análisis de citas; Indicadores bibliométricos; Google Scholar